\documentclass[iop]{emulateapj}

\usepackage[latin1]{inputenc}
\usepackage{amsmath}
\usepackage{natbib}
\usepackage{xcolor}
\usepackage{floatrow}
\usepackage{subfigure}
\usepackage[backref,breaklinks,colorlinks,citecolor=blue,linkcolor=blue]{hyperref} 
\usepackage[all]{hypcap} %Links go to figures;breaks on deluxetables       

\def\co{\mathrm{CO}}
\def\ch{\mathrm{CH}_4}
\def\hh{\mathrm{H}_2}
\def\h2o{\mathrm{H}_2\mathrm{O}}

\def\Ffc{F_{tur}}
\def\kap{K_{zz}}
\def\kaprad{\kappa_{T}}

\def\press{p}
\def\pref{p_0}

\def\ent{s}
\def\Rgas{R}
\def\cp{c_\press}
\def\z{z}
\def\vu{\mathbf{u}}
\def\Tpot{\theta}

\def\Temp{T}
\def\Tempp{T'}

\def\Tempb{T_0}

\def\gradteta{\nabla_{\Tpot}}
\def\gradad{\nabla_{ad}}
\def\gradT{\nabla_{T}}
\def\H{H_p}
\def\ww{w}
\def\wwp{\ww '}

\def\d{\mathrm{d}}

\newcommand{\balign}[1]{
\begin{align}
#1
\end{align}}

\newcommand{\eq}[1]{Equation\,(\ref{#1})}

\newcommand{\fig}[1]{Figure\,\ref{#1}}

\newcommand{\sect}[1]{Section\,\ref{#1}}

\newcommand{\dd}[2]{\frac{\partial #1}{\partial #2}}

\usepackage{setspace}

\begin{document}

\title{Why compositional convection cannot explain \\Substellar objects sharp spectral type transitions.}
%\title{Comment on: "CLOUDLESS ATMOSPHERES FOR L/T DWARFS AND EXTRASOLAR GIANT PLANETS"}

\author{J\'er\'emy Leconte}
\affil{ Laboratoire d'astrophysique de Bordeaux, Univ. Bordeaux, CNRS, B18N, allée Geoffroy Saint-Hilaire, 33615 Pessac, France.}
\email{jeremy.leconte@u-bordeaux.fr}

% Here is to double the space
%\doublespacing
% delete the above line if not double space

\begin{abstract}
As brown dwarfs and young giant planets cool down, they are known to experience various chemical transitions --- for example from $\co$ rich L-dwarfs to methane rich T-dwarfs. Those chemical transitions are accompanied by spectral transitions whose sharpness cannot be explained by chemistry alone. In a series of articles, Tremblin et al. proposed that some of the yet unexplained features associated to these transitions could be explained by a reduction of the thermal gradient near the photosphere. To explain, in turn, this more isothermal profile, they invoke the presence of an instability analogous to fingering convection -- compositional convection -- triggered by the change in mean molecular weight of the gas due to the chemical transitions mentioned above. In this short note, we use existing arguments to demonstrate that any turbulent transport, if present, would in fact \textit{increase} the thermal gradient. This misinterpretation comes from the fact that turbulence mixes/homogenizes entropy (potential temperature) instead of temperature. So, while increasing transport, turbulence in an initially stratified atmosphere actually carries energy downward, whether it is due to fingering or any other type of compositional convection. These processes therefore cannot explain the features observed along the aforementioned transitions by reducing the thermal gradient in the atmosphere of substellar objects. Understanding the microphysical and dynamical properties of clouds at these transitions thus probably remains our best way forward. 
\end{abstract}

\section{Compositional convection}

When the density of a fluid depends on at least two components -- e.g. temperature and composition -- a gradient of composition can trigger turbulent mixing in an otherwise thermally stably stratified medium, a phenomenon that we will call \textit{compositional convection} or \textit{mixing}. If the overall buoyancy gradient is negative, this takes the form of the usual overturning convection \citep{Led47}. If not, some other processes, such as chemistry or diffusion, can still lead to subtle instabilities that enhance mixing. One of the well-known examples here on Earth is the \textit{fingering} instability. For example, when warm salty water resulting from an intense evaporation at the surface of the ocean overlays colder fresh water, sinking \textit{salt fingers} form \citep{Ste60,Sch01}. Although initially buoyant, these downward-moving fingers lose their heat trough diffusion faster than they do their salt and keep sinking \citep{Ste60}. The collective effect of these salt fingers is to increase the turbulent transport in the medium, mixing salt and thermal energy \citep{TSG11}. 

In substellar atmospheres, the range of temperatures encountered entails that various parts of the atmosphere may have very different chemical composition if mixing is not too efficient \citep{ZM14}. Considering carbon chemistry, for example, the deeper/hotter parts of the atmosphere should be dominated by $\co$ and the higher/colder parts by $\ch$ following the net reaction
\balign{\co+3 \hh \rightleftarrows \ch + \h2o.}
This progressive transition from hot $\co$ dominated atmospheres to cold $\ch$ dominated ones is the well known L-T transition (see \citet{Kir05} and \citet{Cus14} for a review). What is more difficult to understand, is both the sharpness of this transition and the fact that its location changes in a color magnitude diagram for various classes of objects (for example high-gravity brown dwarfs versus low-gravity directly imaged planets; \citealt{MSC12}). 

Clouds of various species have long been, and still are, one of the simplest explanations for these various features, although these models still involve some free parameters \citep{CBB17}. In an attempt at reducing the number of these free parameters, \citet{TAC16} proposed a cloud-free model. They noticed that in any single atmosphere around the L-T transition, for example, the chemical equilibrium entails that the colder upper atmosphere should be methane rich and have a higher mean molecular weight compared to the carbon monoxide-rich gas below. \citet{TAC16} thus argued that compositional convection analogous to fingering but linked to chemistry should occur in some brown dwarfs (and young giant planets; \citealt{TCB17}). But for this to explain the observations -- for example an attenuation of the flux in the J band of the objects considered -- mixing would have to decrease the thermal gradient toward the isotherm in the unstable region near the photosphere \citep{TAM15,TAC16,TCB17}. At fixed effective temperature, this indeed causes lower temperatures at depth and lower fluxes in transparent windows (especially the J band). 

It is not clear, however, how they made the link between the presence of compositional convection and the reduction of the thermal gradient. While no demonstration is given, \citet{TAC16}  propose \textit{"that small-scale "diffusive" turbulence, more efficient than radiative transport, induced by fingering convection [...] would be responsible for the decrease of the temperature gradient."}

Such an analogy with radiation seems to imply that the turbulent flux carried by fingering convection, $\Ffc$, could write
\balign{\Ffc=- \rho \cp \kap  \dd{\Temp}{\z},\label{diffflux}}
where $\rho$, $\Temp$, and $\cp$ are the density, temperature, and specific heat capacity of the gas, $\z$ the vertical coordinate, and $\kap$ would be an effective turbulent diffusivity, also known as eddy mixing coefficient. This eventually implies that any turbulence would enhance $\kap$ and thus lead to a stronger upward energy flux that would tend to reduce the thermal gradient toward an isothermal state. 

As will be demonstrated hereafter, this analogy between turbulent and radiative diffusion is not appropriate in this context, even if the turbulence is small-scale. Indeed, as can already be seen from the case of the salt fingers in the ocean, fingering convection does increase the turbulent transport, but carries energy \textit{downward}: The hotter finger from above sinks into a colder fluid and the very reason for the instability is that this finger keeps giving energy to its environment while sinking to remain negatively buoyant. More generally, turbulent mixing in a thermally stably stratified atmosphere leads to entropy mixing and thus to a more adiabatic thermal gradient \citep{Tay15,YM10}.

In the following we argue this using a simple mixing argument in \sect{sec:mixing} and a consideration of the Boussinesq hydrodynamical equations in \sect{sec:downflux}. In \sect{sec:energy}, we briefly discuss why energetic considerations do not preclude a downward energy flux in compositional convection. Let us note that these arguments are not new and can be found in relatively old studies on turbulent transport. Our motivation for briefly rederivating some of these demonstrations here is thus just to gather the necessary pieces for the reader to form an opinion. We conclude that compositional convection cannot explain, through its effects on the thermal gradient, the observed properties of brown dwarfs and directly imaged planets.

\begin{figure*}      % use "figure*" instead of "figure" if you want your figure to span both columns
\epsscale{1.}      % adjust this number to change the size of your figure
%\plotone{Figures/schematicPlot.pdf}
 \hspace*{3cm}\includegraphics[scale=0.2,trim = 1.10cm 6.5cm 8.cm 4.6cm, clip]{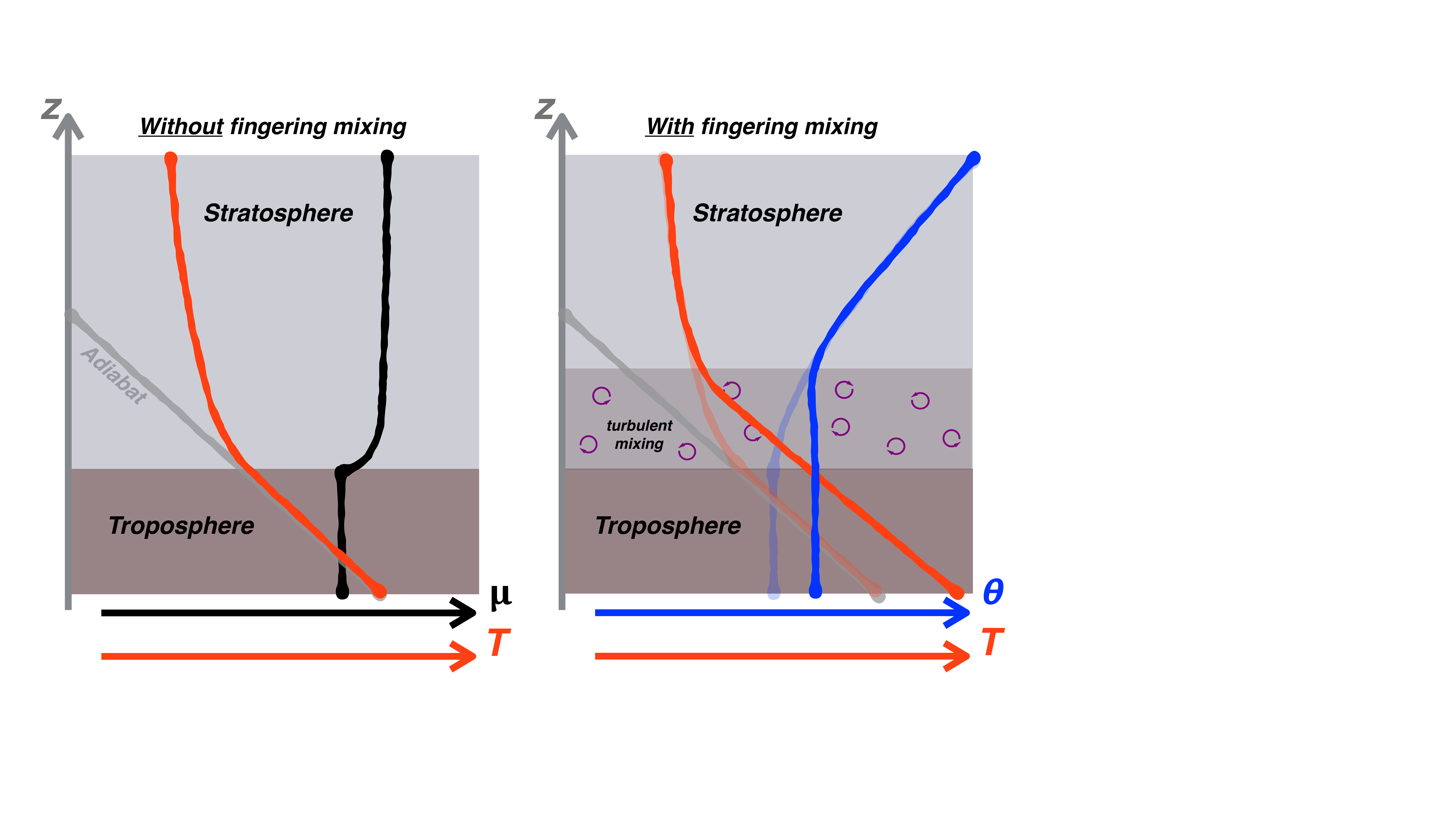}
\caption{A schematic plot showing the thermal (red) and compositional (black) profile in the atmosphere of a brown dwarf in the conventional scenario (left panel). The gray line shows the adiabatic thermal profile which is followed in the troposphere. In the stratosphere, the thermal profile is by definition thermally stably stratified (subadiabatic). However, near the $\co$/$\ch$ the mean molecular weight of the gaz is expected to increase above the tropopause (black curve). Right panel: Modification of the profile of sensible (red) and potential (blue) temperature due to turbulent mixing. The profiles with (without) mixing are the curves with a lighter (darker) shading.  If the compositional gradient is sufficient to trigger compositional convection in the stratosphere, the mixing will homogenize the composition and the entropy (or equivalently the potential temperature; blue curve). This naturally brings the thermal profile back toward the adiabat. The thermal profile is thus \textit{less} isothermal. How close to adiabatic will the resulting profile be depends on the strength of the mixing.}
\label{schematic}
\end{figure*} 

\section{A simple mixing argument}\label{sec:mixing}

Because turbulent mixing entails the motion of fluid parcels, it is important to identify the quantities that are conserved and advected along the motion as these are the quantities that will be \textit{mixed} in the intuitive sense, i.e. homegenized. In a compressible gas, parcels moving adiabatically within an atmosphere do not advect internal energy (temperature), but specific entropy $\ent$. For a perfect gas, it is more intuitive to use the potential temperature
\balign{\Tpot\equiv \Temp \left(\frac{\pref}{\press}\right)^{\Rgas/\cp},}
where $\press$ and $\pref$ are the pressure at the current level and at an arbitrary level of reference, and $\Rgas$ is the gas specific constant. The potential temperature is linked to the entropie through $\d \ent = \cp \d \ln \Tpot$ so that $\Tpot$ is also an advected quantity for an adiabatic motion \citep{Tay15,Val06}. 

The gradient of potential temperature is simply linked to the thermal gradient by
\balign{\gradteta\equiv \frac{\d \ln \Tpot}{\d \ln \press}= \frac{\d \ln \Temp}{\d \ln \press}-\frac{\Rgas}{\cp}\equiv\gradT - \gradad, \label{tetagrad}}
where $\gradad$ is the usual adiabatic thermal gradient. The potential temperature gradient is thus simply the superadiabatic gradient. 

By definition, any turbulent mixing will tend to homogenize the entropy, and thus $\Tpot$, until $\gradteta\rightarrow~0$.  As a result, the atmosphere after mixing tends to follow an adiabatic profile ($\gradT=\gradad$).\footnote{In fact, chemical species are brought aloft where they are unstable and react. The energy deposition is analogous to moist convection where latent heat released by vapor condensation slightly changes the adiabat. For the CO/CH$_4$ reaction, our calculations based on the data from \citet{ZM14} yield a maximum reduction of the adiabatic gradient of $\sim$1\% in a solar metallicity atmosphere, which is too small to explain the observed features.} This is exactly how usual convection works: It homogenizes entropy and potential temperature, removing superadiabaticity. 
In a thermally stably stratified atmosphere, $\gradteta$ is negative ($\gradT<\gradad$), but it works the same, and any mixing will tend to restore an adiabatic profile.\footnote{The reader familiar with the oceanic case might be a little confused by this statement. This is because the adiabatic lapse rate in the ocean is orders of magnitude smaller than in the atmosphere -- roughly 0.1-0.2\,K/km \citep{Tal11} to be compared to 9.8\,K/km -- and so relatively close to the isotherm. Nevertheless, the mixing behavior remains the same: Fingering mixing tends to reduce the thermal stratification of the ocean, cooling the upper/hotter layers and heating the lower/colder ones \citep{Sch01}.} Of course, the extent to which the resulting profile will follow the adiabat cannot be determined a priori and will depend on the strength of the mixing.
As is illustrated in \fig{schematic}, the thermal gradient is therefore \textit{not} reduced toward the isotherm by mixing, but increased toward the adiabat. As a result, compositional convection cannot explain a reduced thermal gradient as presented in \citet{TAM15,TAC16,TCB17}. It does just the opposite!

\section{Downward energy flux in mixed thermally stratified atmospheres}\label{sec:downflux}

What might be a little counter-intuitive about compositional convection -- or any type of turbulence -- bringing a thermally stratified atmosphere toward the adiabat, is that it directly entails that energy is transported \textit{downward} \citep{YM10}. Turbulent mixing actually cools the upper layers and heats the deeper ones. 

The shortest way to make this more explicit is to use the following common approximation for the turbulent flux \citep{Tay15}:
\balign{ \Ffc &=-\rho \cp \kap  \dd{\Tpot}{\z}\label{diffpot}.}
With \eq{tetagrad}, this yields 
\balign{ \Ffc=\rho \cp \kap\frac{\theta}{\H} (\gradT - \gradad)\label{diffpot},}
where $\H$ is the pressure scale height. In the thermally stratified case, the right hand side, hence the flux, is negative.

But \eq{diffpot} is only a working approximation. To show this more rigorously, let us follow an argument from \citet{Mal54}. Consider the energy equation for a compressible fluid in the 
Boussinesq approximation \citep{Bou03,SV60,RGT11}:
\balign{
\dd{\Tempp}{t}+ \vu \cdot\nabla \Tempp+\ww\left(\dd{\Tempb}{\z}-\left.\dd{\Tempb}{\z}\right|_{ad}\right)&=\kaprad \nabla^2 \Tempp  ,
}
where $\kaprad$ is the thermal diffusivity and $\vu=(u,v,w)$ is the velocity perturbation about a state at rest. When needed, quantities are separated into an initial/background state (with a 0 subscript) and a linear perturbation (with a prime). After multiplication by $\Tempp$ and some algebraic manipulations using vector identities we get 
\balign{
\frac{1}{2}\dd{\Tempp^2}{t}&-\frac{1}{2}\Tempp^2 \nabla \cdot \vu +\frac{1}{2}\nabla\cdot  \left(\Tempp^2 \vu\right)-\kaprad\nabla\cdot  \left(\Tempp\nabla \Tempp\right)\nonumber\\
&= -\kaprad \left|\nabla \Tempp \right|^2-\ww \Tempp \left(\dd{\Tempb}{\z}-\left.\dd{\Tempb}{\z}\right|_{ad}\right),}
where all the important terms have been kept on the right hand side \citep{Mal54}.
The second term on the left hand side disappears because of the non divergence of the velocity field in the Boussinesq approximation. To get rid of the others, we average over a large volume encompassing the unstable region in the vertical and with an arbitrary extension in the horizontal (denoted by an overbar).\footnote{Thanks to the Gauss-Ostrogradsky theorem, for any of the terms that write $\nabla\cdot  \mathbf f$ we get 
\balign{\overline{\nabla\cdot  \mathbf f}=\frac{1}{V} \int \nabla\cdot  \mathbf f \,d V=\frac{1}{V} \int_{S_{t},S_{b}}  \mathbf f \cdot \mathbf{d S} +\frac{1}{V} \int_{S_{l}}  \mathbf f \cdot \mathbf{d S}, }
where $S_{t}$, $S_{b}$ and $S_{l}$ are the area of the top, bottom and lateral boundaries of the volume, respectively. The first term on the right hand side vanishes because the top and bottom boundaries are taken outside the turbulent zone, where the perturbations are zero by construction. Because $\mathbf{f}$ is a bounded function, the second term can be made vanishingly small by increasing the volume horizontally while keeping its vertical extent constant.} Finally, when the unstable region has reached a statistical steady state, $\partial \overline{\Tempp^2}/\partial t =0$.

This leaves us with
 \balign{
\overline{\Ffc}\propto \overline{\ww \Tempp} &=-\kaprad\frac{ \overline{\left|\nabla \Tempp \right|^2}}{\left(\dd{\Tempb}{\z}-\left.\dd{\Tempb}{\z}\right|_{ad}\right)}\nonumber\\
&=\ \ \frac{\kaprad \H}{\Tempb}\frac{ \overline{\left|\nabla \Tempp \right|^2}}{\left(\gradT-\gradad\right)}.}
Because $\overline{\left|\nabla \Tempp \right|^2}$ is by construction a definite-positive quantity, it directly results that the sign of the turbulent energy flux is the same as the sign of the superadiabaticity:
\begin{itemize}
\item In a region unstable to usual convection, $\left. \gradT-\gradad >0\right.$ and the turbulent flux is upward to remove the superadiabaticity,
\item In a thermally stably stratified region, $\left.\gradT-\gradad<0\right.$ and the turbulent flux is downward, as advertised.
\end{itemize}

Note that this argument does not depend in any way on the mechanism producing the mixing. It is thus not surprising to recover a negative energy flux in fully non-linear simulations of the fingering instability as can be seen, for example, in Figure\,2 of \citet{TSG11} or in \citet{BGS13}.\footnote{In the latter article, the negative turbulent flux can be inferred by noticing that their definition of the thermal Nusselt number implies $\mathrm{Nu}_T-1=-\wwp \Tempp$ and that this quantity is positive in all figures.} See also \citet{Gar18}.

\section{Energetic considerations}\label{sec:energy}

What is counter-intuitive about a downward energy flux is that, with usual convection, the upward energy flux is directly linked to an upward buoyancy flux which releases potential gravitational energy --- the very energy source that powers convection. It may thus seem that turbulent mixing of a stably stratified atmospheric column \textit{increases} its potential gravitational energy, and thus cannot occur spontaneously.

In the scenario of \citet{YM10}, this apparent paradox is easily solved by acknowledging that their turbulence is externally forced by atmospheric large-scale winds. The external forcing mechanism is thus providing the extra energy powering the motion. This cannot be the case for a spontaneous process. So what is powering the instability?

In compositional convection, the downward buoyancy flux due to temperature is in fact compensated by an upward buoyancy flux due to the mixing of the top-heavy compositional stratification. This was recognized very early in the case of fingering convection \citep{SAB56,Ste60} but is true whenever the medium is thermally stably stratified and the compositional stratification causes the motion. This is why in \citet{TSG11}, for example, the ratio of the thermal to compositional buoyancy flux is always smaller than one in absolute value. As discussed above, this means that the gravitational energy released by moving high mean molecular weight matter from above is larger than the energy needed to carry cold matter upward. 

\section{Conclusions}

We demonstrated that when a stably stratified atmosphere is subjected to compositional convection, or any kind of turbulent mixing, energy is transported downward and the thermal gradient increases toward the adiabatic one. So, if the chemical gradient were to destabilize the atmosphere of a brown dwarf or a giant planet above the troposphere, this would not lead to a more isothermal profile, as advocated by \citet{TAM15,TAC16,TCB17}. On the contrary, it would increase the thermal gradient, thus yielding hotter interiors for the same effective temperature (see \fig{schematic}). Therefore, reasoning in terms of observables, if we were to follow a spectral sequence along the L/T transition at constant effective temperature similar to the one presented in Figure\,3\,b of \citet{TAC16} for example, the troposphere of the model would become colder and colder as the effect of the increased mixing weakens. This would lead to a J band darkening and a disappearance of the FeH feature along this sequence, which is the opposite of what is seen.

Note that, although we focused on the CO/$\ch$ transition here for sake of concreteness, the effect of compositional convection would be the same whatever the cause of the initial mean molecular weight gradient. The above thus applies to all the other chemical transitions as well.

So it seems that, for the moment, the presence of clouds is needed to interpret the current observed features of spectral transitions among substellar objects in a fully physically consistent way. One thing to keep in mind is that if fingering convection is present in substellar atmospheres, it should still affect the mixing of the chemical species. The effect of this mixing remains to be clarified. But it should be noted that on Earth, while there is a positive gradient of mean molecular weight in the atmosphere due to the gradient of water vapor, no atmospheric process has been unequivocally linked to the occurence of fingering because other sources of turbulence and large scale advection dominate. Considering the level of turbulence driven, for example, by overshooting and gravity waves predicted near the photosphere of substellar objects \citep{FAL10}, this statement may apply to these objects as well. 

\acknowledgements
The author wishes to thank T. Guillot and F. Selsis for suggesting and helping with the calculation of the change of the adiabat due to latent energy release, an anonymous referee for suggesting a way to generalize the argument presented in \sect{sec:downflux} to the non-linear regime, and G. Chabrier and P. Tremblin for comments on the initial manuscript. This project has received funding from the European Research Council (ERC) under the European Union's Horizon 2020 research and innovation program (grant agreement No. 679030/WHIPLASH).

\bibliographystyle{apj}
%\bibliography{../biblio}

\end{document}